\documentclass[twocolumn,showpacs,preprintnumbers,amsmath,amssymb]{revtex4}
\input psfig.sty
\def \be {\begin{equation}}

\def \ee {\end{equation}}
\def \bea {\begin{eqnarray}}
\def \eea {\end{eqnarray}}

\usepackage{graphicx}
\usepackage{dcolumn}
\usepackage{bm}

\begin{document}

\title{Clustering, Angular Size and Dark Energy}

\author{R. C. Santos} \email{cliviars@astro.iag.usp.br}

\author{J. A. S. Lima} \email{limajas@astro.iag.usp.br}
\vskip 0.5cm
\affiliation{Departamento de Astronomia, Universidade de S\~ao Paulo, 05508-900 S\~ao
Paulo, SP, Brasil}

\pacs{Dark energy, cosmic distance, angular size, inhomogeneities}
\begin{abstract}
The influence of dark matter inhomogeneities on the angular size-redshift
test is investigated for a large class of flat cosmological models
driven by dark energy plus a cold dark matter component (XCDM). The results
are presented in two steps. First, the mass inhomogeneities are
modeled by a generalized Zeldovich-Kantowski-Dyer-Roeder (ZKDR)
distance which is characterized by a smoothness parameter
$\alpha(z)$ and a power index $\gamma$, and, second,  we provide a
statistical analysis to angular size data for a large sample of
milliarcsecond compact radio sources.  
As a general result, we have 
found that the $\alpha$ parameter is totally  unconstrained 
by  this  sample of angular diameter data.
\end{abstract}

\maketitle

\section{Introduction}
\hspace{0.5cm}An impressive convergence of recent astronomical
observations are suggesting that our world behaves like a spatially
flat scenario dominated by cold dark matter (CDM) plus an exotic
component endowed with large negative pressure, usually named dark
energy \cite{perm98,Riess,Ef02}. In the framework of general
relativity, besides the cosmological constant, there are several
candidates for dark energy, among them: a vacuum decaying energy
density, or a time varying $\Lambda(t)$ \cite{OzTa87}, the so-called
``X-matter" \cite{turner97}, a relic scalar field \cite{PR03}, and a
Chaplygin Gas \cite{kamen}. Some recent review articles discussing
the history, interpretations, as well as, the major difficulties of
such candidates have also been published in the last few years
\cite{review}.

In the case of X-matter, for instance, the dark energy component is
simply described by an equation of state $p_x = \omega\rho_x$. The
case $\omega = -1$ reduces to the cosmological constant,  and
together the CDM defines the scenario usually referred to as
``cosmic concordance model" ($\Lambda$CDM). The imposition $\omega
\geq -1$ is physically motivated by the classical fluid description
\cite{HElis82}. However, as discussed by several authors, such an
imposition introduces a strong bias in the parameter determination
from observational data. In order to take into account this
difficulty, superquintessence or phantom dark energy cosmologies
have been recently considered where such a condition is relaxed
\cite{faraoni02}. In contrast to the usual quintessence model, a
decoupled phantom component presents an anomalous evolutionary
behavior. For instance, the existence of future curvature
singularities, a growth of the energy density with the expansion, or
even the possibility of a rip-off of the structure of matter at all
scales are theoretically expected (\cite{GonFrei} for a
thermodynamic discussion). Although possessing such strange
features, the phantom behavior is theoretically allowed by some
kinetically scalar field driven cosmology \cite{chi00}, as well as,
by brane world models \cite{Shani03}, and, perhaps, more important
to the present work, a PhantomCDM cosmology provides a better fit to
type Ia Supernovae observations than does the $\Lambda$CDM model
\cite{ChouPadm}. Many others observational and theoretical
properties phantom driven cosmologies (more generally, of XCDM scenarios) 
have been successfully
confronted to standard results (see, for instance
\cite{Alc04,estatist,Kasai,L88,Schneider}).

In this context, one of the most important tasks for cosmologists
nowadays is to confront different cosmological scenarios driven by
cold dark matter (CDM) plus a given dark energy candidate with the
available observational data. As widely known, a key quantity for
some cosmological tests is the angular distance-redshift relation,
$D_{A}(z)$, which for a homogeneous and isotropic background, can
readily be derived by using the Einstein field equations for the
Friedmann-Robertson-Walker (FRW) geometry. From $D_{A}(z)$ one
obtains the expression for the angular diameter $\theta (z)$ which can be
compared with the available data for
different samples of astronomical objects \cite{G2004}.

Nevertheless, the real Universe is not perfectly homogeneous, with
light beams experiencing mass inhomogeneities along their way.
Actually, from small to intermediate scales ($\leq 100$Mpc), there
is a lot of structure in form of voids, clumps and clusters which is
probed by the propagating light \cite{Ogu07}. Since the perturbed
metric is unknown, an interesting possibility to account for such an
effect is to introduce the smoothness parameter $\alpha$ which is a
phenomenological representation of the magnification effects
experienced by the light beam. From general grounds, one expects a
redshift dependence of $\alpha$ since the degree of smoothness for
the pressureless matter is supposed to be a time varying quantity
\cite{Kasai,L88}. When $\alpha = 1$ (filled beam), the homogeneous FRW
case is fully recovered; $\alpha < 1$ stands for a defocusing effect
while $\alpha = 0$ represents a totally clumped universe (empty
beam). The distance relation that takes these mass inhomogeneities
into account was discussed by Zeldovich \cite{Zeld} followed by
Kantowski \cite{Kant}, although a  clear-cut application for
cosmology was given only in 1972 by Dyer and Roeder \cite{Dyer}. 
Later on, by considering a perturbed Friedmannian model 
Tomita \cite{tomita98} performed N-body simulations with 
the CDM spectrum in order to determine the distribution for $\alpha$ 
(see also Ref. \cite{tomita99} for a more general analysis involving distances in perturbed models). Many
references may also be found in the textbook by Schneider, Ehlers and
Falco, \cite{SEF}, as well as, in Kantowski \cite{K98,K00,K03}.  

Many studies involving
the ZKDR distances in dark energy models have been published in the
literature. 
Analytical expressions for a general background in the
empty beam approximation ($\alpha = 0$) were derived by Sereno {\it
et al.} \cite{SPS01}. By assuming that both dominant components may
be clustered they also discussed how the critical redshift, i.e., the
value of $z$ for which $D_{A}(z)$ is a maximum (or $\Theta(z)$
minimum), and compared to the homogeneous background results as
given by Lima and Alcaniz \cite{ALDA00}, and, further discussed by
Lewis and Ibata \cite{Lew02}, and Ara\'ujo and Stoeger \cite{Ara07}.
More recently, Demianski {\it et al.} \cite{Dem03}, derived an
useful analytical approximate solution for a clumped concordance
model ($\Lambda$CDM) valid on the interval $0 \leq z \leq 10$.
Additional studies on this subject involving time delay
(Lewis and Ibata \cite{Lew02}; Giovi and Amendola \cite{Gio01}), gravitational
lensing (Kochanek; Kochanek and Schechter \cite{koc02}) or even
accelerated models driven by particle creation  have also been
considered \cite{CdS04,campo}.

Although carefully investigated in many of their theoretical and
observational aspects, an overview in the literature shows that a
quantitative analysis on the influence of dark energy in connection
with inhomogeneities present in the observed universe still remains
to be studied.  Analytical expression for a general applied for the
$\theta(z)$ statistics with basis on a $\Lambda$CDM cosmology with
constant $\alpha$ \cite{AL041}. It was concluded that the best fit
model occurs at $\Omega_M = 0.2$ and $\alpha = 0.8$ whether the
characteristic angular size $\l$ of the compact radio sources is
marginalized. More recently, the smoothness $\alpha$ parameter
was constrained through a statistical
analysis  involving Supernovae Ia data \cite{SLC07}. A $\chi^{2}$-analysis based on the
182 SNe Ia data of Riess {\it et al.} \cite{Riess} constrained the pair of parameters
($\Omega_M,\alpha$) to be $\Omega_M= 0.33^{+0.09}_{-0.07}$ and $\alpha\geq 0.42$ ($2\sigma$).
Such an analysis has also been carried out in the framework of a $\Lambda$CDM cosmology.

In this paper, we focus our attention on X-matter cosmologies with
special emphasis to phantom models ($\omega < -1$) by taking into
account the presence of a clustered cold dark matter. The mass
inhomogeneities will be described by the ZKDR distance characterized
by a smoothness parameter $\alpha (z)$ which depends on a positive
power index $\gamma$. The main objective is to provide a statistical
analysis to angular size data from a large sample of milliarcsecond
compact radio sources \cite{G99} distributed over a wide range of redshifts
($0.011 \leq z \leq 4.72$) whose distance is defined by the ZKDR
equation. As an extra bonus, it will be shown that a pure CDM model
($\Omega_M = 1$) is not compatible with these data even for the
empty beam approximation ($\alpha = 0$).

The manuscript is organized
as follows. In section 2 we outline the derivation of the 
ZKDR equation for a X-CDM cosmology. We also
provide some arguments (see Appendix) for a locally 
nonhomogeneous Universe where
the homogeneous contribution of the dark matter obeys 
the relation $\rho_h = \alpha \rho_{o}(\rho_M/\rho_{o})^{\gamma}$ where $\gamma$
is a positive number, $\rho_M$ is the average matter density and $\rho_o$ its 
present value. In section 3 we analyze the constraints on the
free parameters $\alpha$ and $\Omega_M$ from angular size data. 
We end the paper
by summarizing the main results in section 4.

\section{The Extended ZKDR Equation}

Let us now consider a flat FRW geometry ($c=1$)
\begin{equation}
ds^2 \ = \ dt^2 \ - \ R^2(t)\left(dr^2 + r^2d\theta^{2} +
r^{2}\sin^{2}\theta\,d\phi^{2}\right),
\end{equation}
where $R(t)$ is the scale factor. Such a spacetime is supported by
the pressureless CDM fluid plus a X-matter component of densities
$\rho_M$ and $\rho_x$, respectively. Hence, the total energy
momentum tensor, $T^{\mu\nu} = {T^{\mu\nu}}_{({M})} +
{T^{\mu\nu}}_{({x})}$, can be written as
\begin{equation}\label{EMT}
T^{\mu\nu} = [\rho_{M} + (1 + \omega)\rho_{x}] U^{\mu}U^{\nu} -
\omega\rho_{x} g^{\mu\nu},
\end{equation}
where $U^{\mu}=\delta^{\mu}_o$ is the hydrodynamics 4-velocity of
the comoving volume elements. In this framework, the independent components of the Einstein Field Equations (EFE)
\begin{equation}\label{EFE}
G^{\mu\nu}\equiv R^{\mu\nu} - \frac{1}{2}g^{\mu\nu}R = 8\pi
GT^{\mu\nu},
\end{equation}
take the following forms:
\begin{equation}\label{FRW1}
({\dot{R} \over R})^{2} = H_{o}^{2}\left[\Omega_{\rm{M}}({R_{o}
\over R})^{3} + \Omega_x({R_{o} \over R})^{3(1 + \omega)}\right] ,
\end{equation}
\begin{equation}\label{FRW2}
{\ddot{R} \over R} = -{1 \over
2}H_{o}^{2}\left[\Omega_{\rm{M}}({R_{o} \over R})^{3} + (3\omega +
1)\Omega_x({R_{o} \over R})^{3(1 + \omega)}\right] ,
\end{equation}
where an overdot denotes derivative with respect to time and $H_{o}
= 100h {\rm{Kms^{-1}Mpc^{-1}}}$  is the Hubble parameter. By the
flat condition, $\Omega_x = 1-\Omega_{\rm{M}}$, is the present day
dark energy density parameter. As one may check from (2)-(5), the
case $\omega = - 1$ describes effectively the favored ``cosmic
concordance model" ($\Lambda$CDM).

On the other hand, in the framework of a  comformally flat FRW
metric, the optical scalar equation in the geometric optics
approximation reads (Optical shear neglected) \cite{Sachs61}

\begin{eqnarray}\label{sachs}
{\sqrt{A}}'' +\frac{1}{2}R_{\mu \nu}k^{\mu}k^{\nu} \sqrt{A}=0,
\end{eqnarray}
where $A$ is the beam cross sectional area, plicas means derivative
with respect to the affine parameter describing the null geodesics,
and $k^{\mu}$ is a 4-vector tangent to the photon trajectory whose
divergence determines the optical scalar expansion
\cite{Kasai,Gio01,SPS01}. The circular frequency of the light ray as
seen by the observer with 4-velocity $U^{\alpha}$ is $\omega=
U^{\alpha}k_{\alpha}$, while the angular diameter distance, $D_A$,
is proportional to $\sqrt A$ \cite{SEF}.

As widely known, there is no an acceptable averaging procedure for
smoothing out local inhomogeneities. After Dyer and Roeder
\cite{Dyer}, it is usual to introduce a phenomenological parameter,
$\alpha (z)=1-{\rho_{cl}\over <\rho_M>}$, called the ``smoothness"
parameter. For each value of $z$, such a parameter quantifies  the
portion of matter in clumps ($\rho_{cl}$) relative to the amount of
background matter which is uniformly distributed ($\rho_M$). As a
matter of fact, such authors examined only the case for constant
$\alpha$, however, the basic consequence of the structure formation
process is that it must be a function of the redshift. Combining
equations (2), (3) and (6), after a straightforward but lengthy
algebra one finds that the angular diameter distance, $D_{A}(z)$,
obeys the following differential equation
\begin{equation}\label{angdiamalpha}
 \left( 1+z\right) ^{2}{\cal{F}}
\frac{d^{2}D_A}{dz^{2}} + \left( 1+z\right) {\cal{G}}
\frac{dD_A}{dz} + {\cal{H}} D_A=0,
\end{equation}
which satisfies the boundary conditions:
\begin{equation}
\left\{
\begin{array}{c}
D_A\left( 0\right) =0, \\
\\
\frac{dD_A}{dz}|_{0}=1.
\end{array}
\right.
\end{equation}
The functions ${\cal{F}}$, ${\cal{G}}$ and ${\cal{H}}$ in equation
(7) read
\begin{eqnarray}
{\cal{F}}& =& \Omega_M (1+z)^3 + (1-\Omega_M)(1+z)^{3(\omega +1
)}\nonumber \\ \nonumber \\ {\cal{G}} &=& \frac{7}{2}\Omega_M
(1+z)^3 +\frac{3\omega +7}{2} (1-\Omega_M )(1+z)^{3(\omega
+1)}\nonumber \\ \nonumber
\\ {\cal{H}} &=& \frac{3\alpha(z)}{2}\Omega_M (1+z)^{3} +
\nonumber\\  && +\frac{3(\omega+1)}{2} (1-\Omega_M)(1+z)^{3(\omega +
1)}.
\end{eqnarray}
The smoothness parameter $\alpha(z)$, appearing in the expression of
${\cal{H}}$, assumes the form below (see Appendix A for a detailed
discussion)
\begin{equation}\label{alpha}
\alpha (z) = \frac{\beta_o(1+z)^{3\gamma}}{1 +
\beta_o(1+z)^{3\gamma}},
\end{equation}
where $\beta_o$ and $\gamma$ are constants. Note that the fraction
$\alpha_o = \beta_o/(1 + \beta_o)$ is the present day value of
$\alpha (z)$. In Fig. 1 we show the general behavior of $\alpha(z)$
for some selected values of $\beta_o$ and $\gamma$.

\begin{figure}[t]
\vspace{.2in}
\centerline{\psfig{figure=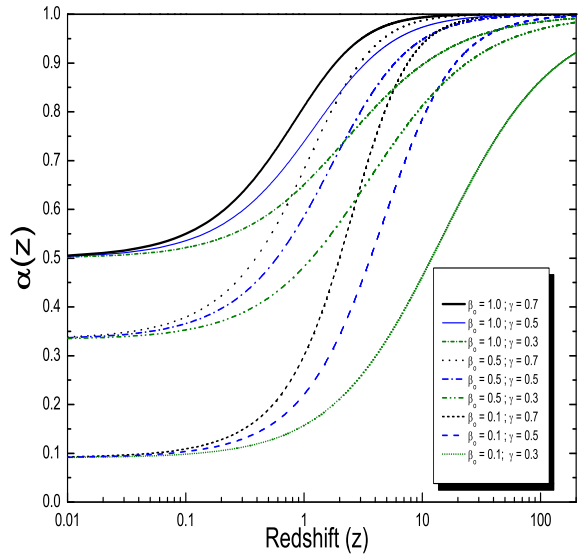,width=3.2truein,height=3.2truein}
\hskip 0.1in} \caption{The smoothness parameter as a function of the
redshift for some selected values of $\beta_o$ and $\gamma$. All
curves approach the filled beam result ($\alpha = 1$) at high
redshifts regardless of the values of $\beta_o$ and $\gamma$. Note
that $\beta_o$ determines $\alpha_o = \alpha(z=0)$. For a given
$\beta_o$ the curves starts at the same point but the rate
approaching unit (filled beam) depends on the $\gamma$ parameter.}
\end{figure}

At this point, it is interesting to compare Eq. (7) together the
subsidiary definitions (8)-(10) with  other treatments appearing in
the literature. For $\gamma = 0$ (constant $\alpha$) and $\omega =
-1$ ($\Lambda$CDM) it reduces to Eq. (2) as given by Alcaniz {\it et
al.} \cite{AL041}.  In fact, for $\omega = -1$ the function
${\cal{H}}$ is given by ${\cal{H}} = \frac{3\alpha}{2}\Omega_M
(1+z)^{3}$. Further, recalling the existence of a simple
relation between the luminosity distance, and the
angular-diameter distance (from Etherington principle \cite{ETHER33},
$D_L = (1+z)^2 D_A$), it is easy to see that Eq. (3) of Santos {\it et al}. \cite{SLC07} is recovered. A more general expression for $\Lambda$CDM model (by
including the curvature term) has been derived by Demianski {\it et
al.} \cite{Dem03}. As one may check,  for $\alpha$ constant, by identifying $\omega \equiv
m/3 -1$, our Eq. (7) is exactly Eq.(10) as presented by Giovi and
Amendola \cite{Gio01} in their time delay studies (see also Eq. (2)
of Sereno {\it et al.} \cite{SPS02}). Different from  other approaches appearing in the literature (see for instance, Refs. \cite{tomita98,tomita99}), we stress  that in this paper the $\alpha$ parameter is always smaller than unity. In addition, the $\alpha$ parameter
may also depend on the direction along the line of sight (for a
discussion of such effects see Linder \cite{L88}, Sereno {\it et al.} \cite{SPS02},
Wang \cite{wan99}).

Let us now discuss the integration of the ZKDR equation with
emphasis in the so-called phantom dark energy model ($\omega < -1$).
In what follows, assuming that $\omega$ is a constant, we have
applied for all graphics a simple Runge-Kutta scheme (see, for
instance, the rksuite package from www.netlib.org).

In Figure 2 one can see how the equation of state parameter,
$\omega$, affects the angular diameter distance. For  fixed values
of $\Omega_M = 0.3$, $\beta_o = 0.5$ and $\gamma = 0$, all the
distances increase with the redshift when $\omega$ diminishes and
enters in the phantom regime ($\omega<-1$). For comparison we have
also plotted the case for $\Lambda$CDM cosmology ($\omega=-1$).

\begin{figure}[t]
\vspace{.2in}
\centerline{\psfig{figure=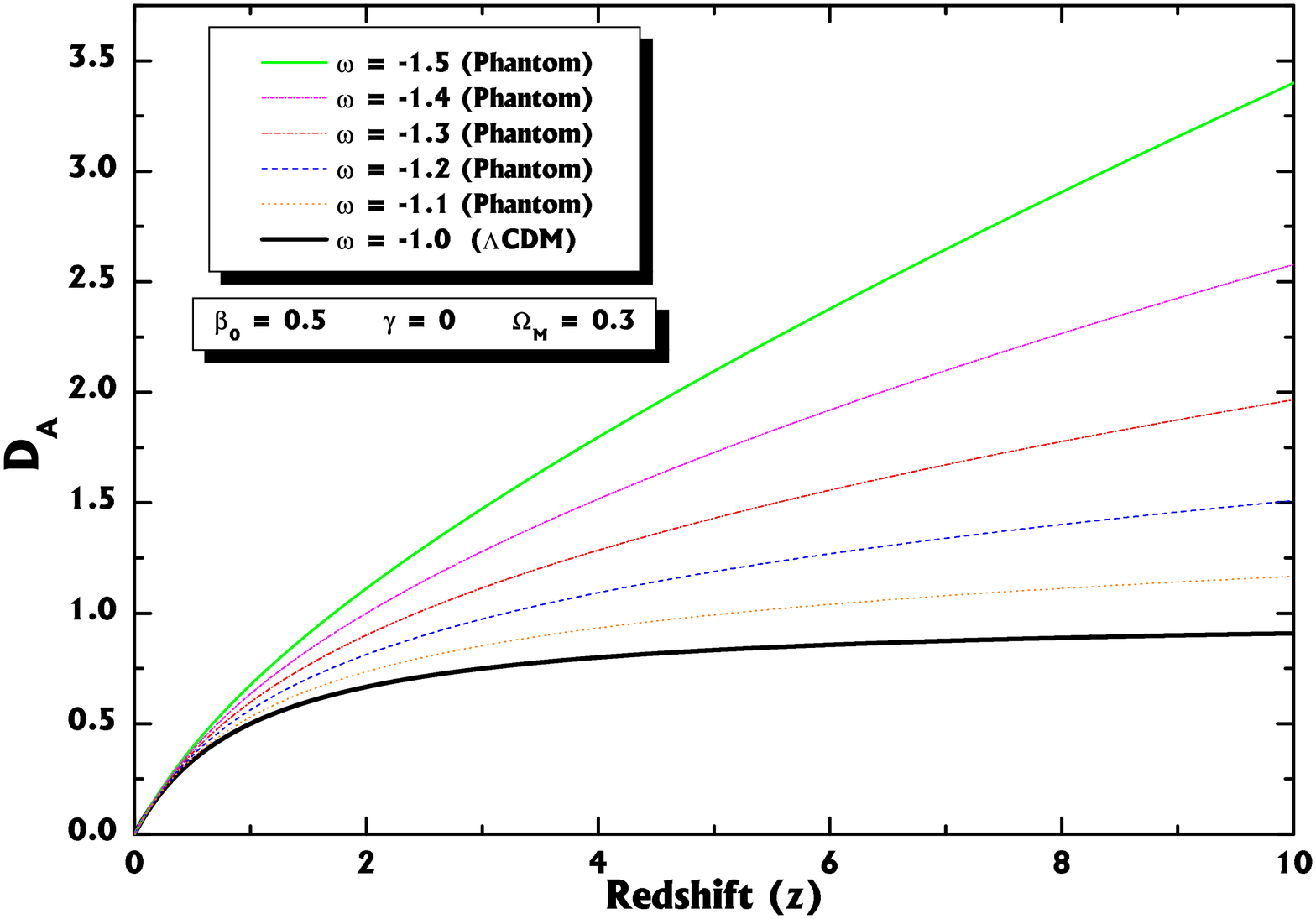,width=3.2truein,height=3.2truein}
\hskip 0.1in}\caption{Angular diameter distance for a flat FRW
phantom cosmology. The curves display the effect of the equation of
state parameter for $\beta_o=0.5$ and $\gamma = 0$. The thick curve
corresponds to the $\Lambda$CDM model. Note that for a given
redshift, the distances always increase for $\omega$ beyond the
phantom divide line ($\omega < -1$).}
\end{figure}

In  Fig. 3 we show the effect of the $\gamma$ parameter on the
angular diameter distance for a specific phantom cosmology with
$\omega = -1.3$, as requested by some recent analyzes of Supernovae
data \cite{Riess}. For this plot we have considered $\beta_o=0.5$.
As shown in Appendix A, $\beta_o = (\rho_h/\rho_{cl})_{z=0}$, is the
present ratio between the homogeneous ($\rho_h$) and the clumped
($\rho_{cl}$) fractions. It was fixed in such a way that $\alpha_o$
assumes the value $0.33$. Until redshifts of the order of 2, the
distance grows for smaller values of $\gamma$, and after that, it
decreases following nearly the same behavior.

In Fig. 4 we display the influence of the $\beta_o$ parameter on the
angular diameter distance for two distinct sets of $\gamma$ values.
The cosmological framework is defined $\Omega_{M} = 0.3$ and the
same equation of state parameter $\omega = -1.3$ (phantom
cosmology). For each branch (a subset of 3 curves with fixed
$\gamma$) the distance increases for smaller values of $\beta_o$, as
should be expected.

\section{ZKDR distance and Angular Size Statistics}

As we have seen, in order to apply the angular diameter distance to
a more realistic description of the universe it is necessary to take
into account local inhomogeneities in  the distribution of matter.
Similarly, such a statement remains true for any cosmological test
involving angular diameter distances,  as for instance, measurements
of angular size, $\theta(z)$, of distant objects. Thus, instead of
the standard FRW homogeneous diameter distance one must consider the
solutions of the ZKDR equation.

\begin{figure}[t]
\vspace{.2in}
\centerline{\psfig{figure=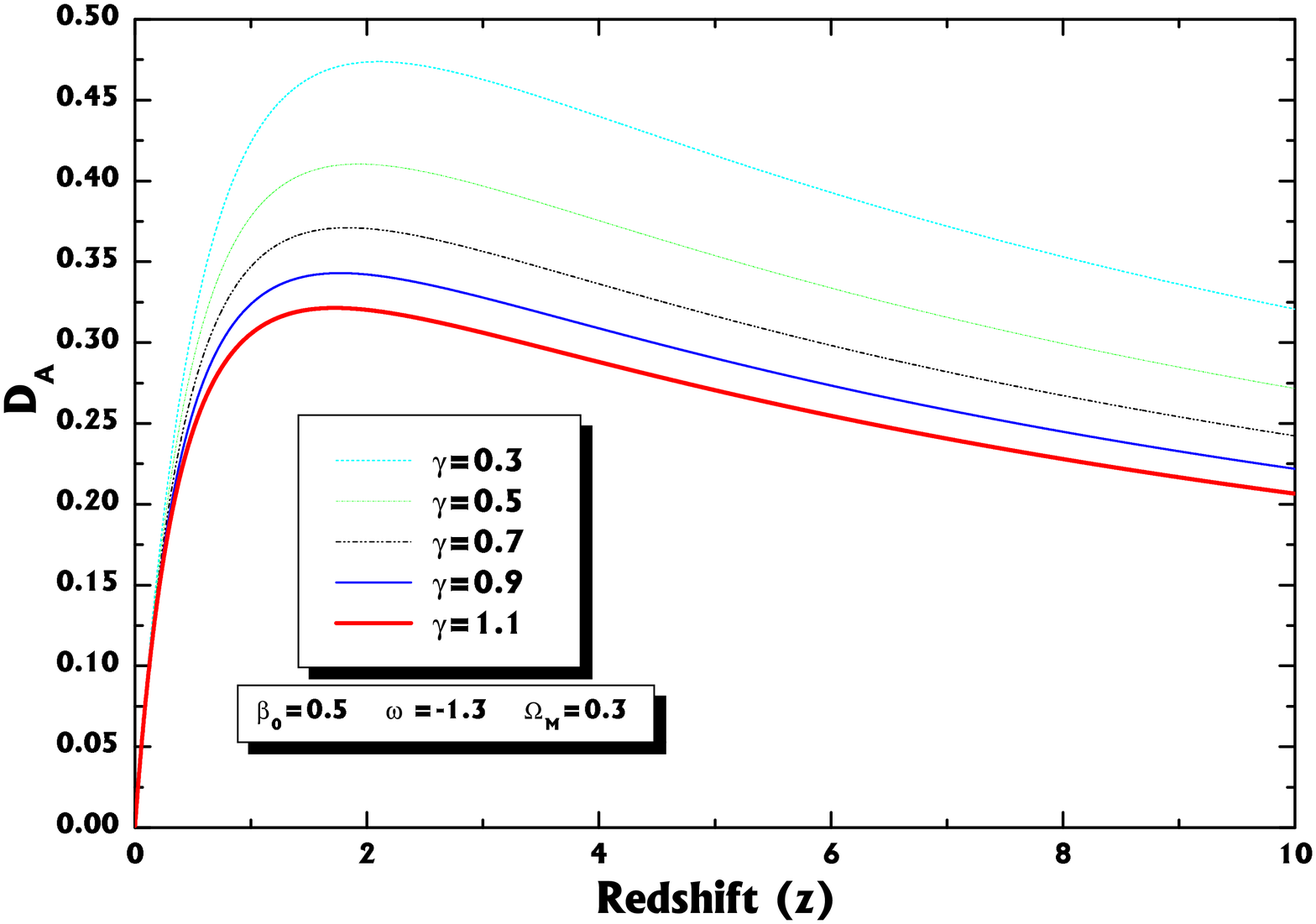,width=3.2truein,height=3.2truein}
\hskip 0.1in} \caption{Effects of the $\gamma$ parameter on the
angular diameter distance. For all curves we fixed $\omega=-1.3$,
$\beta_o = 0.5$ and $\Omega_M=0.3$. Note that the distances increase
for smaller values of $\gamma$.}
\end{figure}

\begin{figure}[t]
\vspace{.2in}
\centerline{\psfig{figure=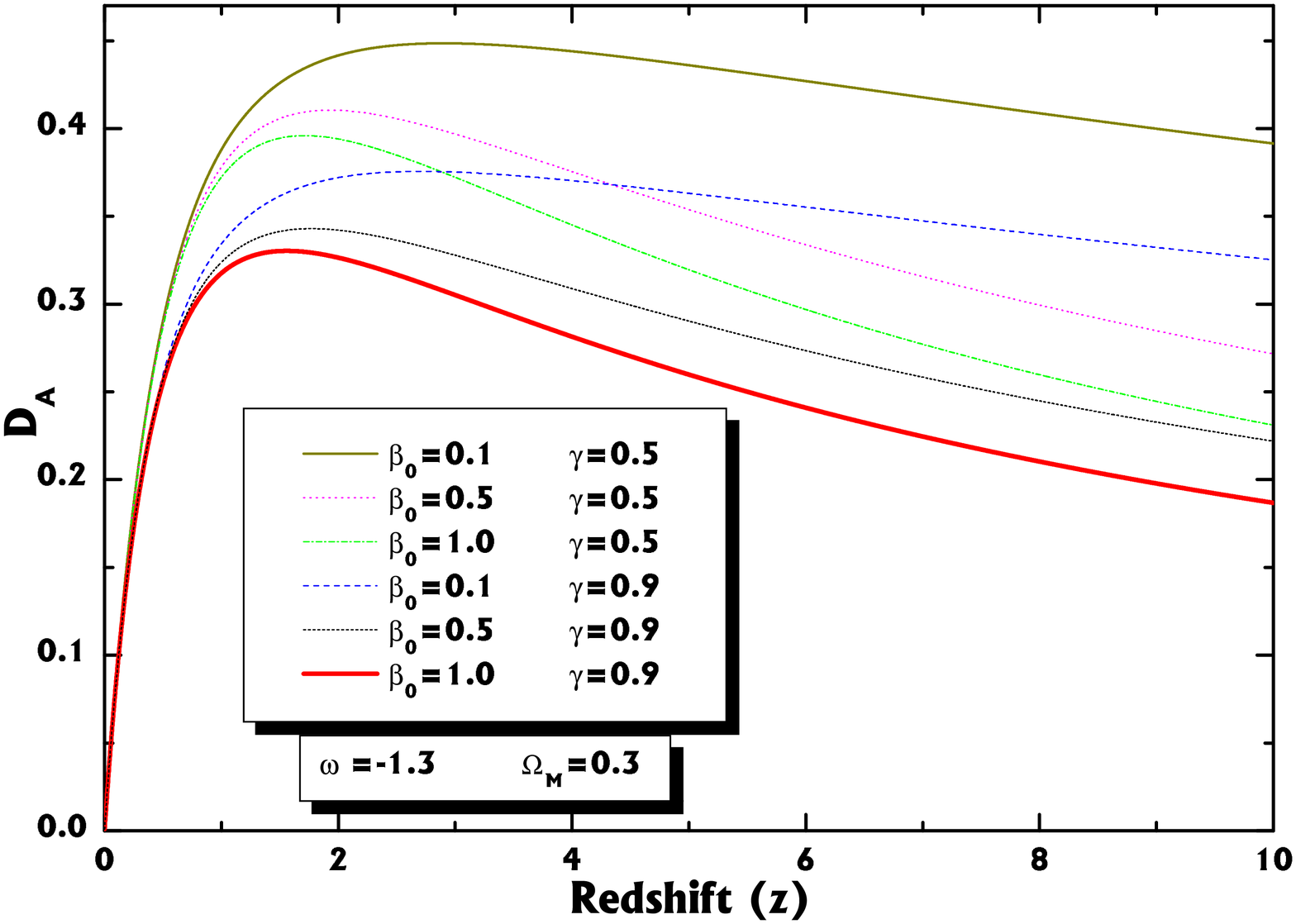,width=3.2truein,height=3.2truein}
\hskip 0.1in} \caption{Influence of the $\beta_o$ parameter on the
angular diameter distance for $\Omega_M=0.3$ and $\omega=-1.3$. The
curves are separated in two sets corresponding to the values of
$\gamma=0.5, 0.9$ as indicated in the box. As expected, both sets
present the same behavior at low redshifts.}
\end{figure}
Here we are concerned with angular diameters of light sources
described as rigid rods and not isophotal diameters. In the FRW
metric, the angular size of a light source of proper length ${\l}$
(assumed free of evolutionary effects) and located at redshift $z$
can be written as
\begin{equation}
\theta(z)= \frac{\ell}{{D_{A}}(z)},
\end{equation}
where $\ell = 100{\l}$h is the angular size scale expressed in
milliarcsecond (mas) while ${\l}$ is measured in parsecs for compact
radio sources (see below).

Let us now discuss the constraints from  angular size measurements
of high $z$ objects on the cosmological parameters. The present
analysis is based on the angular size data for milliarcsecond
compact radio sources compiled by Gurvits {\it et al.} \cite{G99} (see also \cite{G2004} for applications to the unclustered FRW case).
This sample is composed by 145 sources at low and high redshifts
($0.011 \leq z \leq 4.72$) distributed into 12 bins with 12-13
sources per bin (for more details see Gurvits {\it et al.}
\cite{G99}). In Figure 5 we show the binned data of the median
angular size plotted as a function of redshift $z$ to the case with
$\gamma=0$ and some selected values of $\Omega_M$ and $\alpha_o =
\beta_o/(1 - \beta_o)$ = constant. As can be seen there, for a given
value of $\Omega_M$ the corresponding curve is slightly modified for
different values of the smoothness parameter $\alpha$.

Now, in order to constrain the cosmic parameters, we first fix the
central value of the Hubble parameter obtained by the Hubble Space Telescope (HST) key
project $H_o = 72 \pm 8$ ${\rm{km.s^{-1}.Mpc^{-1}}}$ (Freedman {\it
et al.} \cite{F01}). Nowadays, this HST result has been confirmed by many different classes of 
estimators like the Sunyaev-Zeldovich effect and the ages of old high redshifts galaxies \cite{CML}. This value is also in accordance
with the 3 years release of the WMAP team \cite{Ef02}, however, it  is greater than the recent 
determination by Sandage and collaborators \cite{San06}. Following
standard lines, the confidence regions are constructed through a
$\chi^{2}$ minimization
\begin{equation}
\chi^{2}(l, \omega, \alpha) =
\sum_{i=1}^{12}{\frac{\left[\theta(z_{i}, \l, \omega, \alpha) -
\theta_{oi}\right]^{2}}{\sigma_{i}^{2}}},
\end{equation}
where $\theta(z_{i}$, $\l$, $\omega$, $\alpha)$ is defined from Eq.
(7) and $\theta_{oi}$ are the observed values of the angular size
with errors $\sigma_{i}$ of the $i$th bin in the sample. The
confidence regions  are defined by the conventional two-parameters
$\chi^{2}$ levels. In this analysis, the intrinsic length $\l$, is
considered a kind of ``nuisance" parameter, and, as such, we have
also marginalized over it.
\begin{figure}[t]
\vspace{.2in}
\centerline{\psfig{figure=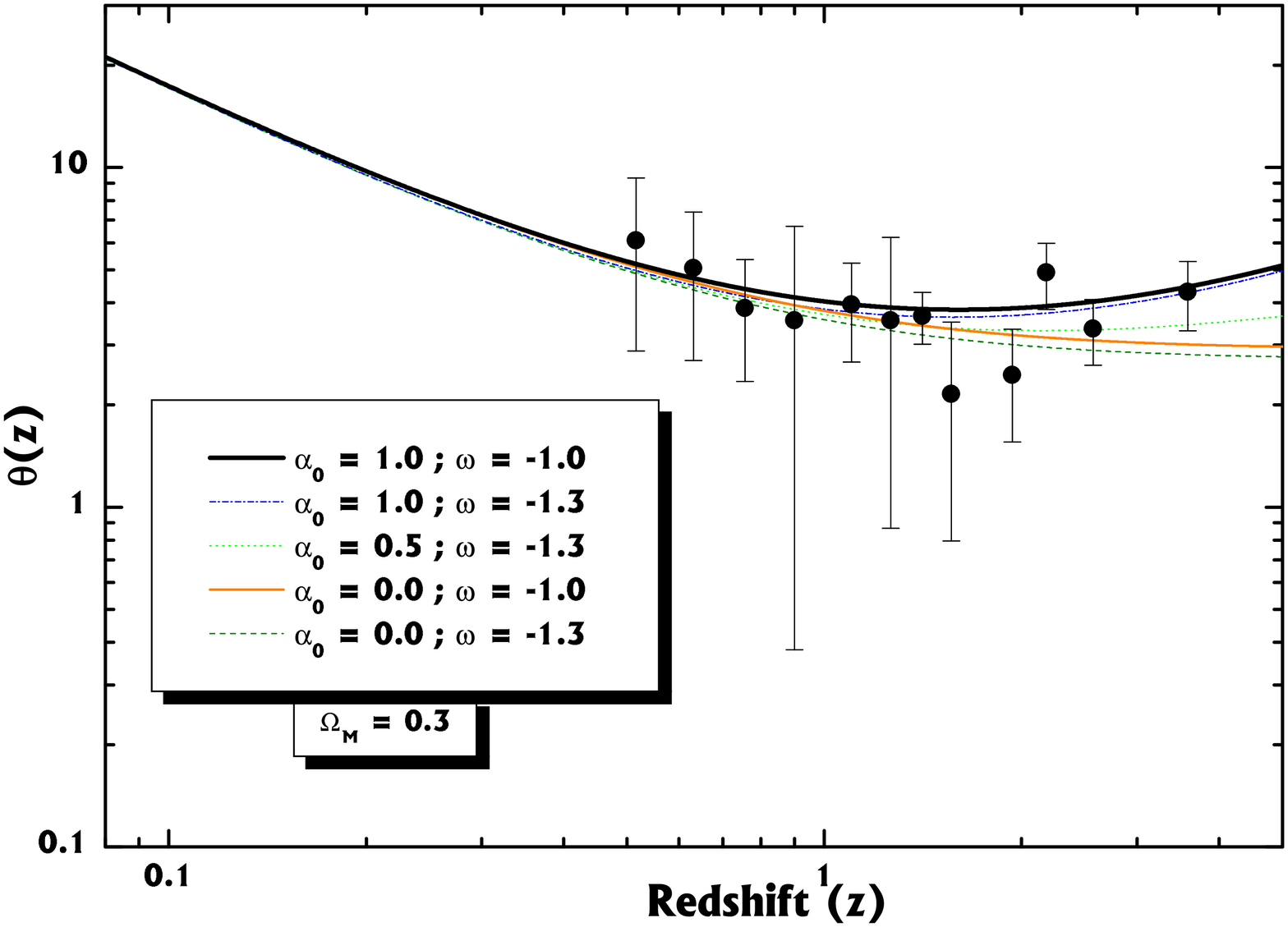,width=3.2truein,height=3.2truein}
\hskip 0.1in} \caption{Angular size versus redshift according to the
ZKDR distance. Curves for $\Omega_M=0.3$, $\gamma=0$ and different
values of $\omega$ are shown. The data points correspond to 145
compact radio sources binned into 12 bins (Gurvits {\it et al.}
\cite{G99}). For comparison the filled beam $\Lambda$CDM has been
included.}
\end{figure}

\begin{figure}[t]
\vspace{.2in}
\centerline{\psfig{figure=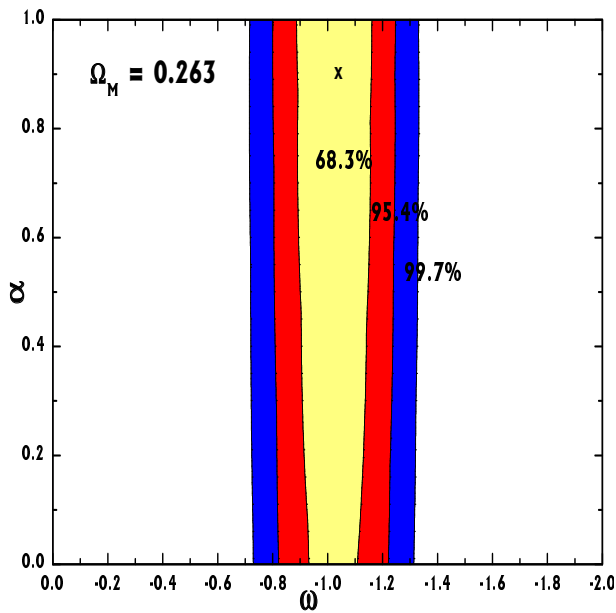,width=3.2truein,height=3.2truein}
\hskip 0.1in} \caption{Confidence regions in the $\omega - \alpha$
plane according to the sample of angular size data by Gurvits {\it
et al.} \cite{G99} and fixed $\Omega_M = 0.263$ as shown in panel.
The confidence levels of the contours are indicated. The point ``x"
marks the best fit values, $\omega = -1.03$ and $\alpha = 0.90$.}
\end{figure}

\begin{figure}[t]
\vspace{.2in}
\centerline{\psfig{figure=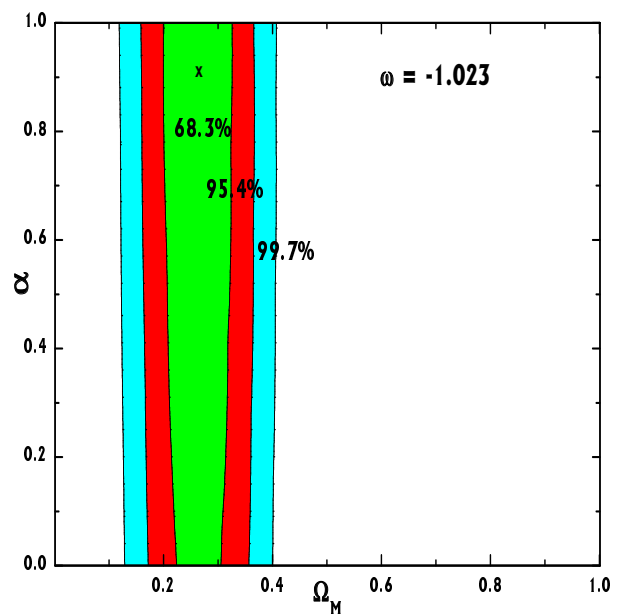,width=3.2truein,height=3.2truein}
\hskip 0.1in} \caption{Confidence regions in the $\Omega_M-\alpha$
plane according to the sample of angular size data by Gurvits {\it
et al.} \cite{G99}. For a phantom cosmology with $\omega=-1.023$,
the confidence levels of the contours are indicated. As in Fig. 6,
the ``x" also points to the best fit values shown in the panel.}
\end{figure}
In Fig. 6 we show confidence regions in the $\omega - \alpha$ plane
fixing $\Omega_{M} = 0.263$, and assuming a Gaussian prior on the
$\omega$ parameter, i.e., $\omega = -1 \pm 0.3$ (in order to
accelerate the universe). The ``$\times$" indicates the best fit
model that occurs at $\omega = -1.03 $ and $\alpha \simeq 0.9$.

In Fig. 7 the confidence regions are shown in the $\Omega_M-\alpha$
plane. We have now assumed a Gaussian prior on $\Omega_{M}$, i.e.,
$\Omega_{M} = 0.3 \pm 0.1$ from the large scale structure. From
Figs. 6 and 7, it is also perceptible that while the parameters
$\omega$ and $\Omega_M$ are strongly restricted, the entire interval
of $\alpha$ is still allowed. This shows the impossibility of
tightly constraining the smoothness parameter $\alpha$ with the
current angular size data. This result is in good agreement with the
one found by Lima and Alcaniz \cite{OzTa87} where the same data set
were used to investigate constraints on quintessence scenarios in
homogeneous background, and is also in line with the one obtained by
Barber et al. \cite{Barb00} who argued in favor of $\alpha_o =
\alpha(z=0)$ near unity (see also Alcaniz, Lima and Silva
\cite{AL041} for constraints on a clustered $\Lambda$CDM model).

\section{Summary and Concluding Remarks}

All cosmological distances must be notably modified whether the
space-time is filled by a smooth dark energy component with negative
pressure plus a clustered dark matter. Here we have addressed the
question of how the angular diameter distance of extragalactic
objects are modified by assuming a slightly inhomogeneous universe.
The present study complements our previous results \cite{G2004} by considering that the
inhomogeneities can be described by the
Zeldovich-Kantowski-Dyer-Roeder distance (in this connection see
also, Giovi and Amendola \cite{Gio01}; Lewis and Ibata \cite{Lew02};
Sereno {\it et al.} \cite{SPS02}; Demianski {\it et al.}
\cite{Dem03}). The dark energy component was described by the
equation of state $p_x = \omega \rho_x$. A special emphasis was
given to the case of phantom cosmology ($\omega < -1$) when the
dominant energy condition is violated. The effects of the local
clustered distribution of dark matter have been described by the
``smoothness" phenomenological parameter $\alpha (z)$, and a simple
argument for its functional redshift dependence was given in the
Appendix A (see also Figure 1).

The influence of the dark energy component was quantified by
considering the angular diameters for sample of milliarcsecond radio
sources (Fig. 5) as described by Gurvits {\it et al.} \cite{G99}.
By marginalizing over the characteristic angular size $l$, fixing $\Omega_M=0.263$, and
assuming a Gaussian prior on the EOS parameter, i.e.,
$\omega = -1 \pm 0.3$, the best fit
model occurs at $\omega =-1.03$ and $\alpha = 0.9$. This phantom
model coincides with the central value recently determined by the
Supernova Legacy Survey (Astier {\it et. al.} \cite{Ef02}). On the
other hand, fixing $\omega = -1.023$ and assuming a Gaussian prior for
$\Omega_{M}$, that is, $\Omega_{M} = 0.3 \pm 0.1$, we obtained the
best fit values ($ \Omega_{M} = 0.29$, $\alpha = 0.9$).

Finally, in order to improve the present
results, a statistical study is necessary for determining the intrinsic length
of the compact radio sources.  Further, unlike to what happens with SNe data \cite{SLC07}, 
the angular diameter sample of compact radio sources of Gurvits et al. 
\cite{G99} does not provide useful constraints on the $\alpha$ parameter (see Figs. 6 e 7). 
Naturally, these results reinforce the interest for measurements of angular size from
compact radio sources at intermediary and high redshifts in order to constrain the $\alpha$ 
parameter with  basis on the ZKDR distance. 
\appendix
\section{On the redshift dependence of $\alpha(z)$}

In this Appendix we discuss the functional redshift dependence of
the smoothness parameter, $\alpha(z)$, adopted in this work. By
definition
\begin{equation}
\alpha(z) = 1 - \frac{\rho_{cl}(z)}{\rho_M(z)},
\end{equation}
where $\rho_{cl}$ denotes the clumped fraction of the total matter
density, $\rho_M$, present in the considered FRW type Universe. This
means that the ratio between the homogeneous ($\rho_h$) and the
clumped fraction can be written as $\rho_h/\rho_{cl} = \alpha(z)/[1-
\alpha(z)]$. How this ratio depends on the redshift? In this concern,
we first remember that $\alpha(z)$ lies on the interval [0,1].
Secondly, in virtue of the structure formation process, one expects
that the degree of homogeneity must increase for higher redshifts,
or equivalently, the clumped fraction should be asymptotically
vanishing at early times, say, for $z \geq 100$. This means that
$\alpha (z) \rightarrow 1$ at high z. On the other hand, $\alpha$ must be zero 
for a completely clustered matter which is disproved at low redshifts by the 
data of galaxy clusters \cite{Ef02}. It thus follows that at present ($z=0$), the related
fraction assume an intermediate value, say, $\beta_o$. In
addition, it is also natural to suppose that the redshift dependence
of the total matter density, $\rho_M$, must play an important role
in the evolution of their fractions. In this way, for the sake of
generality, we will assume a power law
\begin{equation}
\frac{ \rho_h}{\rho_{cl}} \equiv \frac{\alpha(z)}{1- \alpha(z)} =
\beta_o (\frac{\rho_M}{\rho_o})^{\gamma}.
\end{equation}
where $\beta_o = (\rho_h/\rho_{cl})_{z=0}$ and $\gamma$ are
dimensionless numbers. Finally, inserting $\rho_M (z)$, and solving
for $\alpha(z)$ we obtain:
\begin{equation}
\alpha(z) = \frac{\beta_o(1 + z)^{3\gamma}}{1 + \beta_o(1 +
z)^{3\gamma}},
\end{equation}
which is the expression adopted in this work (see Eq. (\ref{alpha})).

As one may check, for positive values of $\gamma$, the smoothness
function (A3) has all the physically desirable properties above
discussed. In particular, the limit for high values of $z$ does not
depend on the values of $\beta_o$ and $\gamma$ (both of the order of
unity). Note also that if the clumped and homogeneous portions are
contributing equally at present ($\beta_o=1$), we see that
$\alpha(z=0) = 1/2$ regardless of the value of $\gamma$. Figure 1
display  the general behavior of $\alpha(z)$ with the redshift for
different choices of $\beta_o$ and $\gamma$. The above functional
dependence should be compared with the other ones discussed in the
literature (see, for instance, \cite{Kasai,L88,campo} and Refs. therein).
One of the most interesting features of (A3) is
that its validity is not restricted to a given redshift interval.

\begin{acknowledgements}
The authors would like to thank A. Guimar\~aes and J. V. Cunha for
helpful discussions. RCS thanks CNPq No. 15.0293/2007-0 and JASL
thanks CNPq and FAPESP grant No. 04/13668.
\end{acknowledgements}

\end{document}